\documentclass{article}

\usepackage{amsmath}
\usepackage{microtype}
\usepackage{graphicx}
\usepackage{subfigure}
\usepackage{booktabs} 

\usepackage{hyperref}

\usepackage[accepted]{icml2019}

\icmltitlerunning{Two-level Explanations in Music Emotion Recognition}

\begin{document}

\twocolumn[
\icmltitle{Two-level Explanations in Music Emotion Recognition}



\icmlsetsymbol{equal}{*}

\begin{icmlauthorlist}
\icmlauthor{Verena Haunschmid}{jku,equal} 
\icmlauthor{Shreyan Chowdhury}{jku,equal}
\icmlauthor{Gerhard Widmer}{jku,lit}
\end{icmlauthorlist}


\icmlaffiliation{jku}{Johannes Kepler University Linz}
\icmlaffiliation{lit}{LIT AI Lab, Linz Institute of Technology (LIT)}

\icmlcorrespondingauthor{Verena Haunschmid}{verena.haunschmid@jku.at}
\icmlcorrespondingauthor{Shreyan Chowdhury}{shreyan.chowdhury@jku.at}

\icmlkeywords{Machine Learning, ICML, Interpretable Machine Learning, Emotion Recognition}

\vskip 0.3in
]



\printAffiliationsAndNotice{\icmlEqualContribution} 

\begin{abstract}
Current ML models for music emotion recognition, while generally working quite well, do not give meaningful or intuitive explanations for their predictions. In this work, we propose a 2-step procedure to arrive at spectrogram-level explanations that connect certain aspects of the audio to interpretable mid-level perceptual features, and these to the actual emotion prediction. That makes it possible to focus on specific musical reasons for a prediction (in terms of perceptual features), and to trace these back to patterns in the audio that can be interpreted visually and acoustically.
\end{abstract}

\section{Introduction}
\label{sec:introduction}

Emotions play an important part in music. Previous work has shown that emotion can be predicted from audio with satisfactory precision using deep neural networks. A limitation of deep models in general is their lack of transparency, which makes it impossible to understand how they arrived at their predictions. To address this problem, we proposed a two-level model in \cite{anonymous} that learns to predict 8 emotion qualities (e.g., valence, tension, sadness)
by first predicting a set of 7 \textit{mid-level perceptual features} \cite{aljanaki2018} from audio, and then predicting the emotions from these, via a linear network layer (see the last two layers in the network model in Fig.\ref{fig:arch}). The linear nature of this last layer permits us to obtain intuitive explanations (in the form of \textit{effects plots}) of predicted emotions in terms of the perceptual features that gave rise to the predictions.

In the present work, to provide more complete explanations, we propose an extension of the model in which listenable explanations for these mid-level features are created, based on local interpretable model-agnostic explanations (LIME)~\cite{ribeiro2016}.
In this way, we close the gap between the prediction of an abstract emotion label and a visual and acoustic presentation of the aspects of the audio that led to the prediction (which may also be a starting point for targeted audio manipulations in some innovative application scenarios).  


\section{Two-level Emotion Recognition}

\begin{figure}
    \centering
    \includegraphics[trim={0 0.3cm 0 0.1cm}, clip, width=\columnwidth]{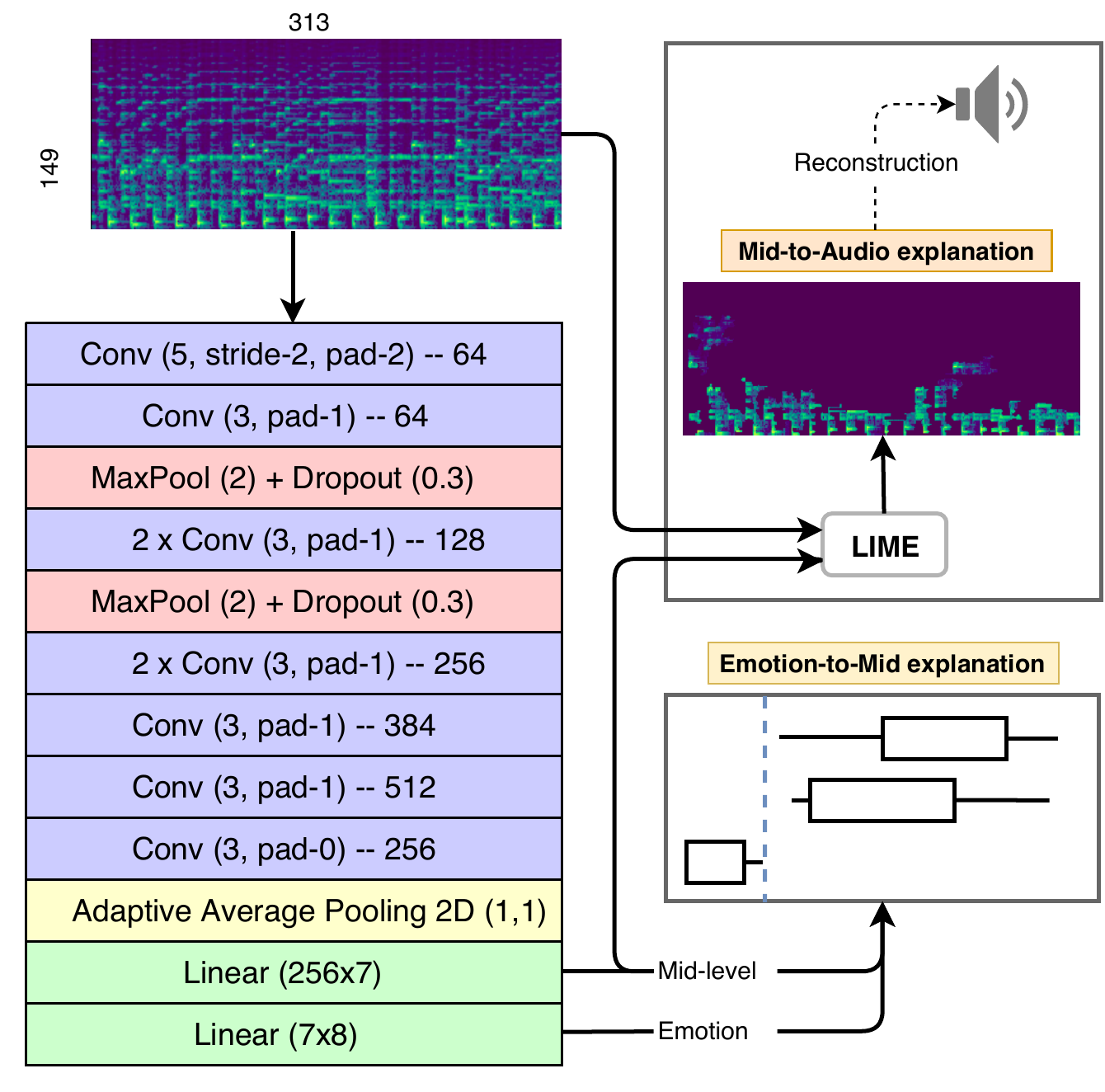}
    \vspace{-0.3cm}
    \caption{Overview of the proposed system. Left: architecture of the audio-to-emotion network. The final two layers are designed to represent mid-level (\textit{Mid}) qualities and emotions and their linear relationship. Lower right: Emotion-to-Mid explanations provided by effects plots. Upper right: the proposed extension for Mid-to-Audio explanations based on a modified version of LIME.}
    \label{fig:arch}
\end{figure}

Our starting point is an audio-to-emotion model introduced in~\cite{anonymous} (termed \emph{A2Mid2E-Joint}). It is based on a VGG-style convolutional neural network with an intermediate layer that models mid-level (\textit{Mid}) features followed by the final (linear) layer modeling emotion (left half of Figure~\ref{fig:arch}). 
Mid-level perceptual features should represent musical
qualities (such as \textit{melodiousness} and \textit{rhythmic complexity}) that are easily perceived and recognized
by most listeners, without music-theoretic training.
By optimizing on a combined loss during training the model learns to predict mid-level features and emotion ratings jointly. The model has been trained on the \textit{Soundtracks dataset} for which both emotion annotations~\cite{eerola2011} (8 emotions) and mid-level perceptual feature annotations~\cite{aljanaki2018} (7 mid-level qualities) are available. The two-level model's performance (in terms of Pearson correlation, averaged over all emotions) has been shown to be virtually equal to that of a simpler, non-interpretable baseline model (without the additional mid-level layer).

\section{Obtaining Two-level Explanations}

\cite{anonymous} showed how explanations can be derived from the trained model by exploiting the linearity of the last (Mid-to-Emotion) 
layer. \textit{Effects plots} (hinted at in the lower right part of Fig.\ref{fig:arch}) display the influence of mid-level features on emotion predictions, both globally for the entire learned model, and for individual predictions.

Inspired by~\cite{mishra2017soundlime}, 
we now show how to use LIME~\cite{ribeiro2016} to get explanations for the Audio-to-Mid part of our emotion predictor. To derive an explanation for an instance LIME trains a simpler, interpretable model on a set of perturbed samples by using the original model prediction as the label. The weights of the simpler model are then used to determine which ``features'' (e.g., image segments) are important to the prediction. Our implementation is based on the original version of LIME\footnote{\url{https://github.com/marcotcr/lime/}}. 





One of the main issues with LIME is the instability of the results when repeating the sampling process~\cite{molnar2019}. Preliminary experiments showed that, unsurprisingly, the instability is related to the number of features $N$ and therefore number of possible perturbed samples ($2^N$). To circumvent this, all experiments have been conducted with a sufficiently large number of sampled instances (50k) which is far from the default value of LIME (1k). Another crucial parameter is the number of important features that should be reported. To avoid manual evaluation, we devise a strategy to automatically select the optimal number of features per instance. We do this by calculating the p-values for each of the weights and thresholding on the 
$\text{p-value}:\text{weight}$
ratio. For our experiments, we observed a ratio of $10^{-6}$ to work well, which selects about 30 to 60 features from a total of about 300.
Finally, regarding possible image segmentation algorithms (Felzenszwalb, SLIC, Chan Vese, Watershed), experiments showed that Felzenszwalb (with $\mathit{scale}=25$ and $\mathit{min\_size}=40$) provided the most reasonable visual segmentation of the spectrograms.

The final output of the Audio-to-Mid explanation process are two spectrograms that show the image segments with positive and negative weights, respectively -- in other words, those aspects of the audio that most strongly contributed to the prediction, in a positive or negative way. The other parts of the spectrogram are hidden.
From this, we can synthesise `enhanced' versions of the original audio which make the relevant musical aspects more clearly audible, by adding (or subtracting) the segments with the original spectrogram, and inverting this modified spectrogram back to time domain.

\section{Example and Discussion}

\begin{figure}[t!]
    \centering
    \includegraphics[width=\columnwidth]{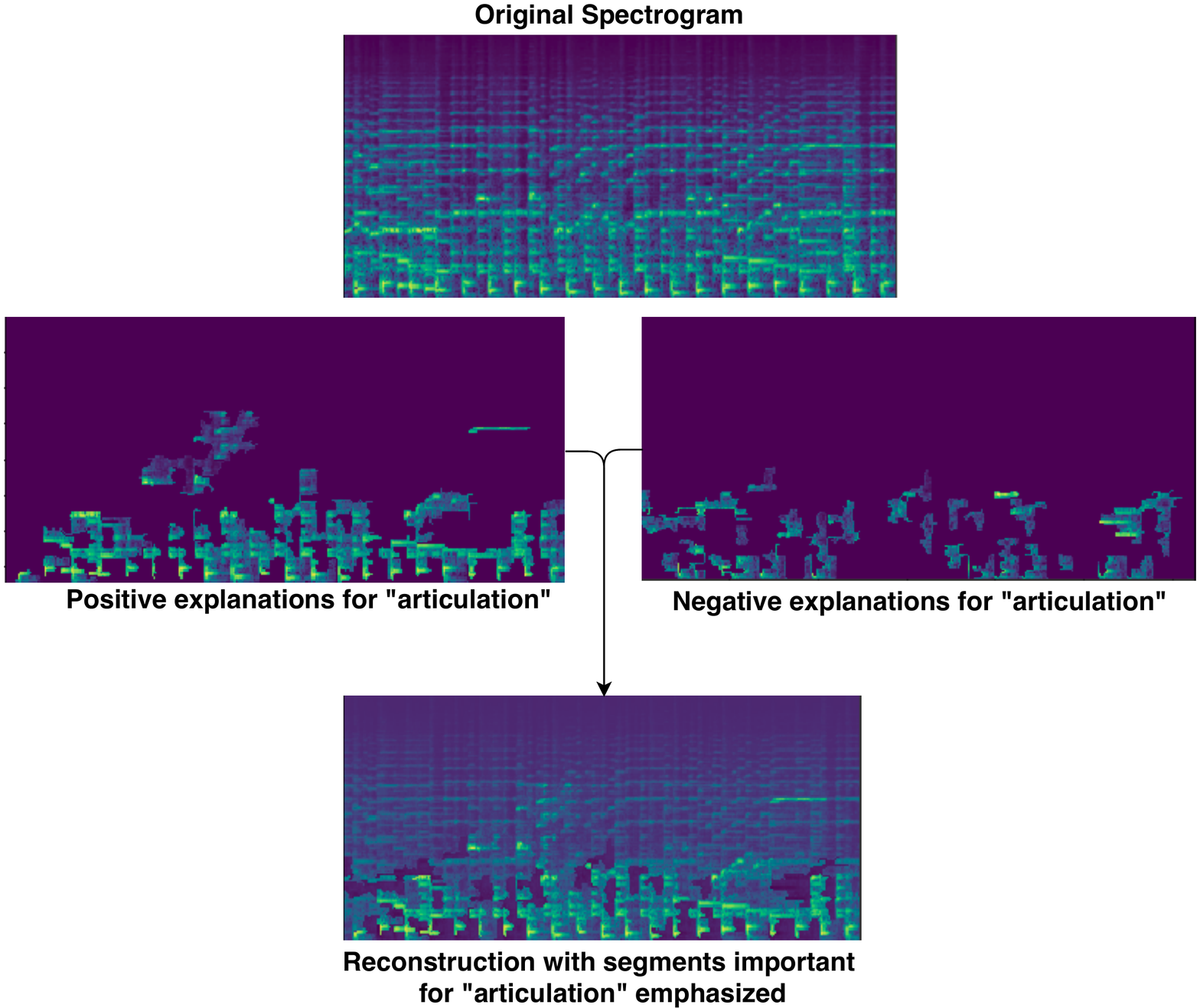}
    \vspace{-0.3cm}
    \caption{Spectrograms visualizing explanations for the mid-level feature ``articulation'' for the song.}
    \label{fig:explanations}
\end{figure}

We demonstrate our method by selecting an arbitrary audio clip (file 001.mp3, 10 sec) from our collection and running our trained model on it. 
(Please go to \url{https://shreyanc.github.io/ICML_example.html} for details and audios.)
The model predicts an emotional profile for the song. The dimension with the highest predicted value is ``energy", which our Mid-to-Emotion explanation (the effects plot) traces back, in particular, to a high perceived ``articulation" (``staccatoness") quality.
This in turn is explained by our Audio-to-Mid explanation tool by pointing to certain patterns in the spectrogram (see also Fig.\ref{fig:explanations}).
Visually, these look intuitively meaningful in relation to a perceived percussive quality. Listening to the re-synthesised ``enhanced" version based on the explanation (``Modified" in the linked web page) makes it clear that it is the drums that strongly contributed to the prediction. This in itself is probably not very useful, as the concept of ``articulation" or ``percussiveness" is already intuitively interpretable to us. However, the audio explanation could be the basis for targeted modifications of the audio, by which we could try to enhance or weaken certain musical qualities in a directed way to control perceived emotional qualities of the music, which might enable some interesting applications. This is an interesting starting point for future work.


\section*{Acknowledgements}

This research has received funding from the European Research Council 
(ERC) under the European Union's Horizon 2020 research and innovation 
programme, grant agreement No 670035 (``Con Espressione'').



\bibliography{library}
\bibliographystyle{icml2019}

\end{document}